\documentclass[reprint,amsmath,aps,pra,footinbib,notitlepage,longbibliography,superscriptaddress]{revtex4-2}
\usepackage{graphicx,epstopdf,dcolumn,bm,mathtools,siunitx,multirow,hyperref,xcolor,amsfonts,physics}

\usepackage{float} 
\usepackage{afterpage}
\usepackage{hyperref} 
\begin{document}
\title{Shape-controlled Bose-Einstein Condensation}

\author{Cem Kurt}
\affiliation{Department of Physics, Koç University, 34450 Sariyer, Istanbul, Turkey}

\author{Altug Sisman}
\affiliation{Department of Physics and Astronomy, Uppsala University, 75120 Uppsala, Sweden}

\author{Alhun Aydin}
\email{alhun.aydin@sabanciuniv.edu}
\affiliation{Faculty of Engineering and Natural Sciences, Sabanci University, 34956 Tuzla, Istanbul, Turkey}
\affiliation{Department of Physics, Harvard University, Cambridge, Massachusetts 02138, USA}

\date{\today}
\begin{abstract}
Size-invariant shape transformation is a geometric technique that allows for a clear separation between quantum size and shape effects by modifying the shape of the confinement domain without altering its size. The impact of shape on the behavior of confined systems is significantly different from that of size, making it an emerging area of research. The recent realization of flat-bottomed optical box traps has further contributed to the study of quantum gases in complex confinement geometries. Here, we propose shape-induced Bose-Einstein condensation at a fixed size, temperature, and density. We investigate the impact of pure quantum shape effects on a non-interacting Bose gas confined within nested square domains, where the shape parameter is defined and controlled by the rotation angle between the inner and outer squares. Our findings reveal that specific heat exhibits an additional low-temperature peak at certain shapes. This work opens new avenues for controlling quantum systems through geometric manipulation and provides insights into the thermodynamic properties of Bose gases under shape-induced quantum effects. 
\end{abstract}
\maketitle


\section{Introduction}

Bose-Einstein condensation (BEC) is a macroscopic quantum phenomenon in which a significant fraction of bosonic particles occupy the ground state~\cite{BECbook}. Traditionally, condensation has been achieved using harmonic traps, with key control parameters such as temperature, particle density, trap frequency, and particle interaction strength~\cite{RevModPhys.71.463,RevModPhys.73.307}. These harmonic electromagnetic traps have been the mainstay of BEC experiments~\cite{RevModPhys.74.875}. Conversely, recent experimental breakthroughs have introduced flat-bottomed optical box traps, allowing for the realization of homogeneous Bose gases~\cite{Gaunt2013,Chomaz2015,Navon2015,Lopes2017,PhysRevLett.118.123401,Navon2021}. This advancement enables the experimental investigation of ultracold atomic gases in flat-bottomed potentials, also known as particle-in-a-box potentials, a feat previously unattainable. Uniform optical box traps pave the way for exploring Bose-Einstein condensates in more complex optical box shapes~\cite{Navon2021,PhysRevResearch.2.043256,doi:10.1126/science.abm2543,Tononi2023,TONONI20241}.

Quantum-confined particles in box-like potentials with well-defined size parameters are often discussed in the context of quantum size effects, where energy quantization becomes prominent as the thermal de Broglie wavelength of particles, $\lambda_{th}=\hbar\sqrt{2\pi}/\sqrt{mk_BT}$, approaches the box size~\cite{bineker,PhysRevLett.120.170601}. The size parameters are defined under the Lebesgue measure as volume, surface area, periphery, and the number of vertices~\cite{pathbook,aydin3}, which are collectively called Weyl size parameters. Both the dimensionality and sizes are critical control parameters for the thermodynamic properties of a Bose gas as well as the BEC transition. Quantum size effects in Bose-Einstein condensates have been explored extensively in harmonic traps~\cite{GROSSMANN1995188,Grossmann1995,HAUGERUD199718,WangJian,PhysRevA.72.063611,PhysRevA.54.4188,PhysRevA.76.063604,PhysRevA.55.2922,PhysRevA.58.1490,Cheng2021,NORONHA2016485}, rotating traps~\cite{RevModPhys.81.647}, systems with finite number of particles~\cite{PhysRevA.54.656,PhysRevA.55.3954,PhysRevA.104.043318,PhysRevA.83.023616,Brange2023}, low-dimensional traps\cite{PhysRevLett.87.130402,Petrov2000,PhysRevA.59.4657,sym13020300}, and anisometric traps with multistep condensations~\cite{PhysRevLett.79.549,PhysRevD.60.105016,multi1,multi2}. However, a more intriguing scenario emerges when the size of the confining potential remains constant, but its shape is altered at fixed temperature and density.

\begin{figure}
    \centering
    \includegraphics[width=1\linewidth]{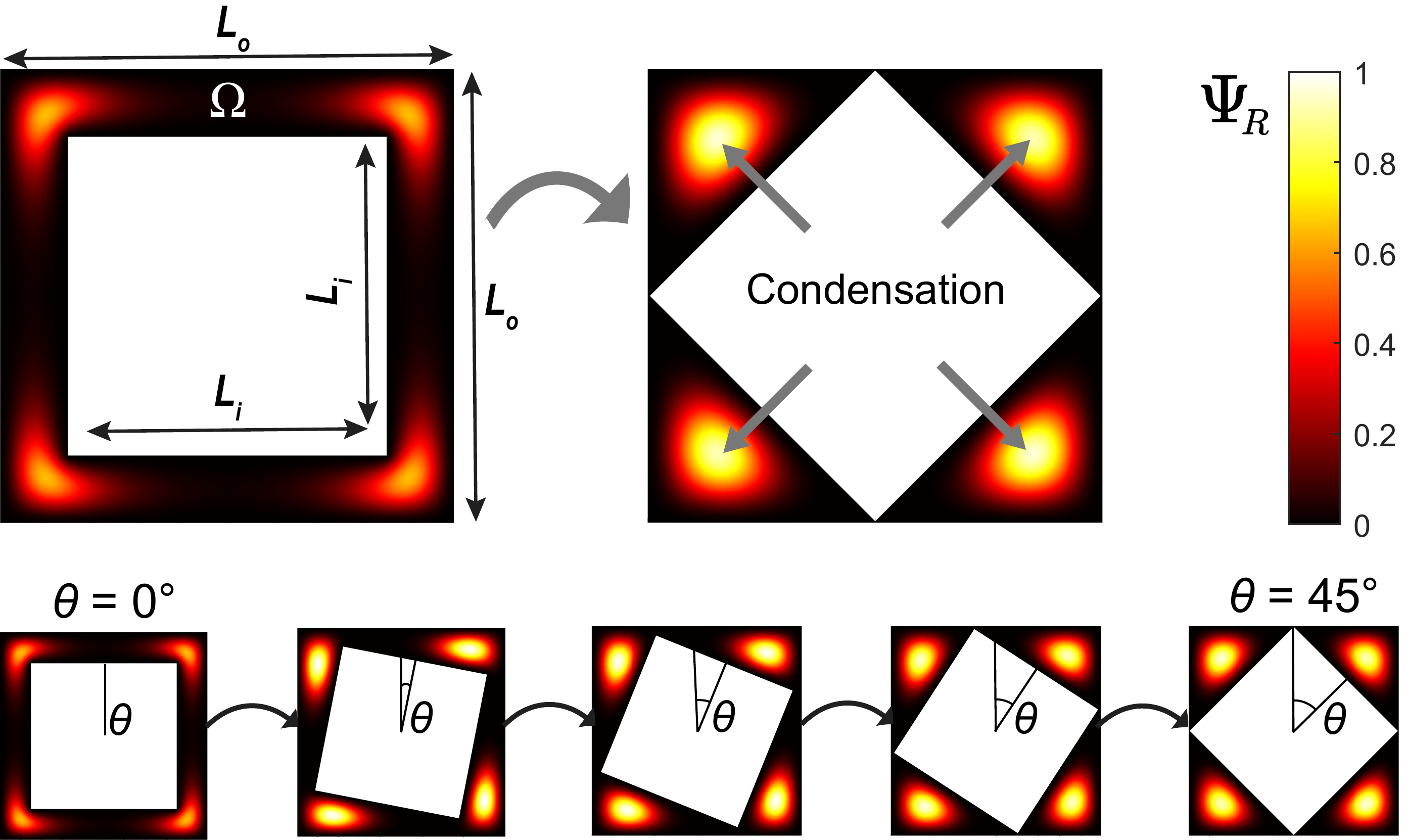}
     \caption{Bose-Einstein condensation induced by the size-invariant shape transformation. A Bose gas is confined in a core-shell nested square domain, $\Omega$. The shape parameter, $\theta$, is defined as the angle between the inner and outer squares. The vertex distance is defined as the perpendicular distance from a vertex of the inner square to the nearest side of the outer square. The normalized equilibrium probability density distributions of the ground state ($\Psi_R=|\psi_{GS}(\theta)|^2/|\psi_{GS}(45^{\circ})|^2$) for different configuration angles are shown at a specific temperature and density. With the quasistatic rotation of the inner square, the degenerate ground state becomes macroscopically occupied despite the sizes of the domain remaining constant.}
     \label{fig:figure1}
\end{figure}

While previous studies have investigated BEC in various arbitrarily shaped potentials~\cite{ZIFF1977169,PhysRevE.59.158,PhysRevA.94.053609,PhysRevA.104.063310,PhysRevLett.129.243402,OZTAS2024129853,Herbst2024}, these cannot be considered as pure shape effects, because they do not ensure size-invariance; inevitably, one or more size parameters change, making the size effects unavoidably vary. Conversely, a novel quantum-mechanical phenomenon known as the quantum shape effect occurs when the geometry of the box potential is modified while keeping all the Lebesgue size parameters, as well as the topology and boundary curvature of the confinement domain, unchanged~\cite{Aydin2019,aydinphd,Aydin2023}. The quantum shape effect is achieved through a geometric technique called size-invariant shape transformation~\cite{specsist}, illustrated in Fig. 1. 

Quantum shape effects are distinct from quantum size effects, as they can induce fundamentally different and unexpected behaviors~\cite{Aydin2023}. For instance, quantum shape effects can lead to unusual thermodynamic phenomena, such as spontaneous transitions to lower entropy states with significant implications for quantum thermodynamics~\cite{specsist} and quantum energy devices\cite{aydin11}. In fermionic systems, quantum shape effects can result in shape-dependent quantum oscillations, offering a novel approach to band-gap engineering~\cite{Aydin2022}. These unique properties underscore the critical importance of investigating quantum shape effects in the context of BEC. Understanding how the quantum shape effect influences BEC transitions could open new avenues for controlling quantum systems.

In this study, we investigate the condensation of a non-interacting Bose gas with a finite number of particles confined in a 2D flat-bottom potential between nested square domains with impenetrable boundaries. The angle between the sides of the inner and outer squares defines a size-invariant shape parameter. We demonstrate that quasistatic changes in a shape parameter can induce the BEC. The condensate fraction can be changed and controlled just by shape, without altering the size parameters of the confinement domain, as well as the density and temperature of the gas. Then, we investigate the behavior of Bose-Einstein gas under different dimensionless temperatures and densities to find the optimum condition where the change in condensate fraction due to shape effects is maximized. We calculate the condensate fraction and heat capacity of a Bose gas changing with shape, temperature, and density. Note that our focus is solely on demonstrating that shape alone can induce the condensation of a Bose gas. As long as the overall confinement is sufficiently strong, the effects of geometry can be made dominant over inter-particle interactions~\cite{PhysRevE.70.016103,Pang_2006,10.1063/1.2821248,aydinphd}. Consequently, we do not employ the Gross-Pitaevskii equation, as the dynamics or other properties of the condensate are beyond the scope of this study.

\begin{figure*}[t]
    \centering
    \includegraphics[width=\textwidth]{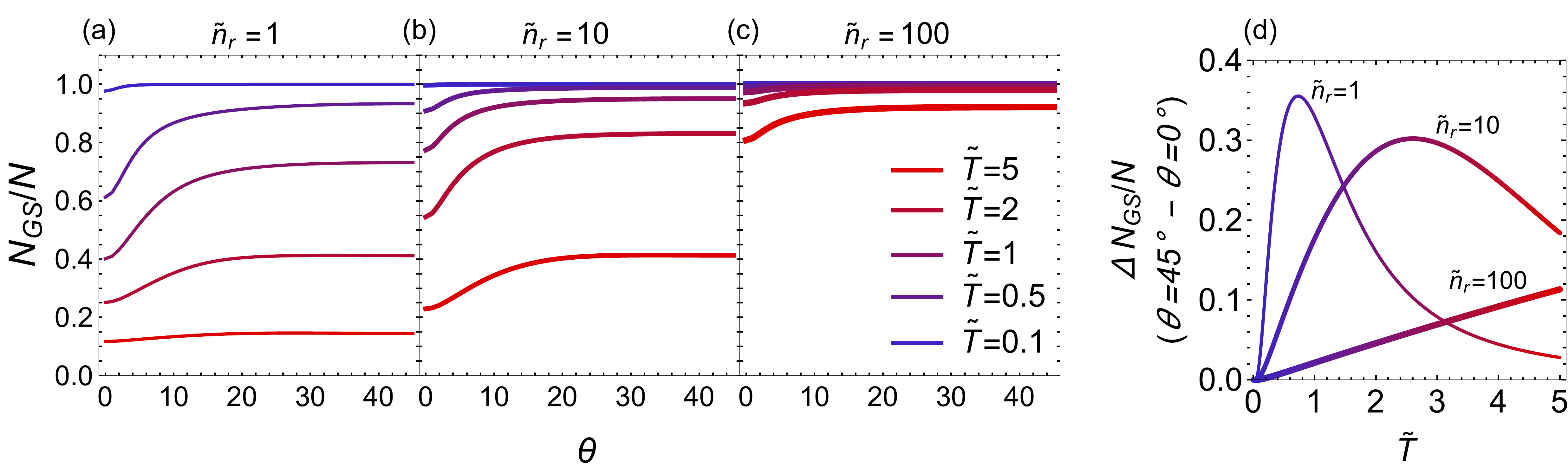}
    \caption{Shape dependence of the condensate fraction for different temperatures at densities (a) $\tilde{n}_r=1$, (b) $\tilde{n}_r=10$, (c) $\tilde{n}_r=100$. (d) Difference of condensate fraction between $\theta=45^{\circ}$ and $\theta=0^{\circ}$ configurations as a function of $\tilde{T}$ for different $\tilde{n}_r$.}
    \label{fig:figure2}
\end{figure*}

\begin{figure}
\centering
\includegraphics[width=1\linewidth]{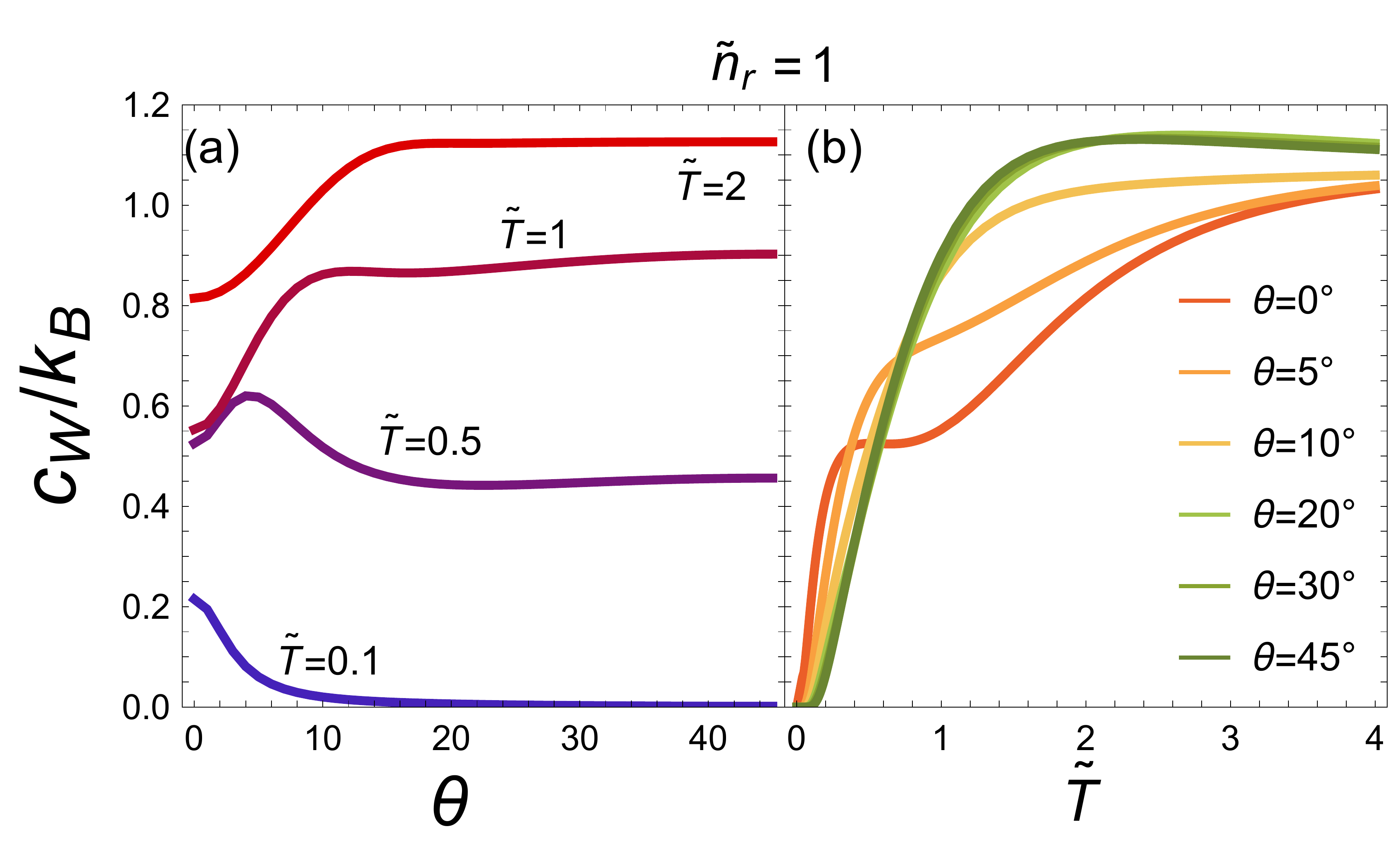}
\caption{(a) Shape dependence of the specific heat at constant Weyl size parameters for $\tilde{n}_r = 1$ in the units of $k_B$. (b) Variation of the specific heat with respect to temperature for various angular configurations.}
\label{figure3}
\end{figure}

\begin{figure*}
\centering
\includegraphics[width=\linewidth]{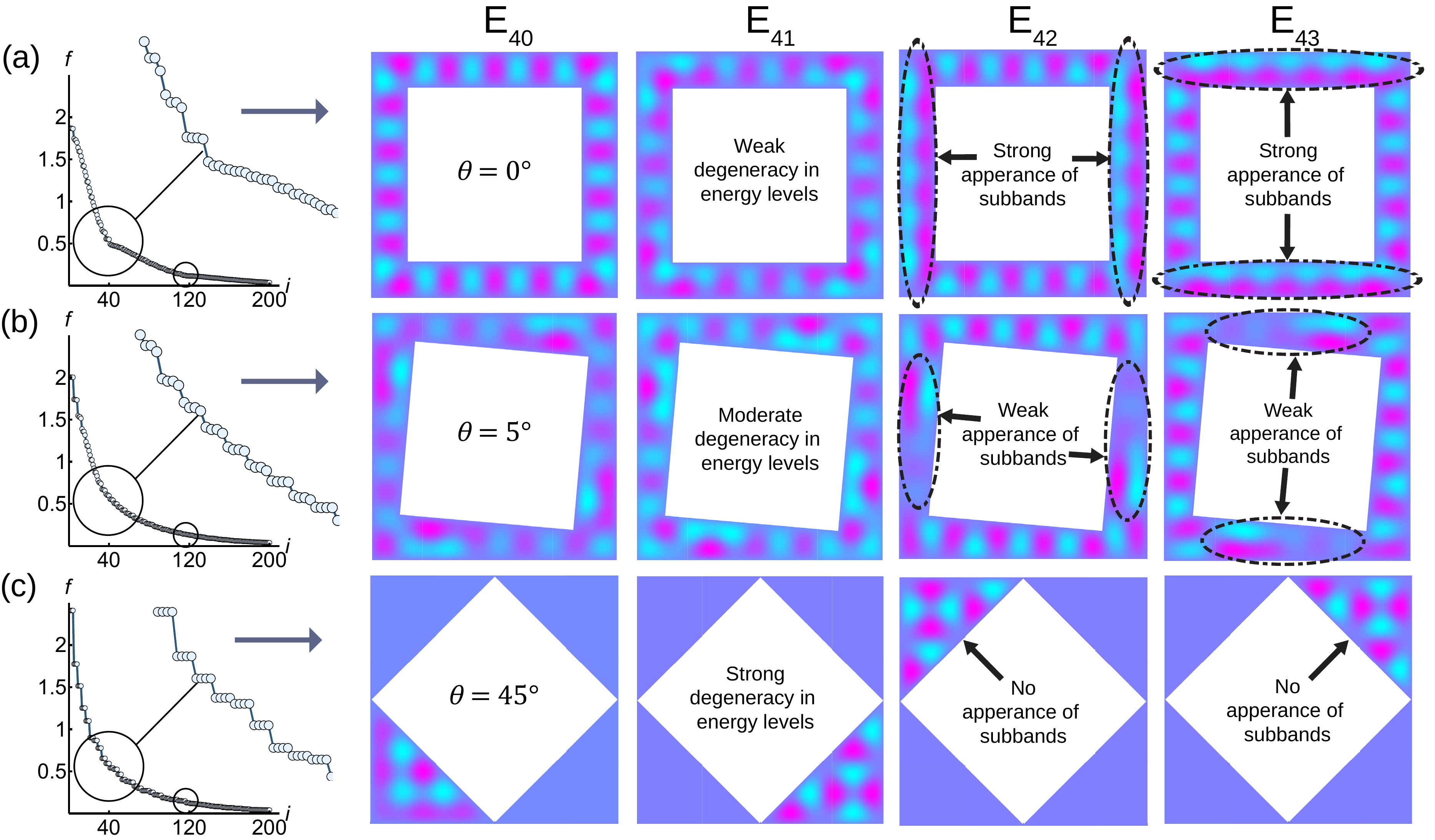}
\caption{Level-degeneracy and local subband behaviors due to the quantum shape effect. On the left column, thermal occupancies of quantum states, denoted by the momentum state variable $i$, are shown for (a) $\theta = 0^{\circ}$, (b) $\theta = 5^{\circ}$ and (c) $\theta = 45^{\circ}$. Eigenfunctions of four consecutive states are plotted next to them for the region where the first subband appears. As the $\theta$ increases the strength of subband excitation gradually disappears and fourfold degeneracy takes over.}
\label{fig:figure4}
\end{figure*}

\section{Confinement geometry}
A non-interacting Bose gas is confined in the region between two nested square domains as shown in Fig. 1 at $\tilde{T} =1$ and $\tilde{n}_r =1$, with the potential being infinite inside the inner square and outside the outer square, and zero in between. The outer boundaries of the domain are defined by the side length $L_o=L$, while the inner boundaries are set to $L_i=0.7L$. The shape control is governed by the parameter $\theta$, which represents the rotation angle of the inner core, allowing us to make size-invariant changes to the domain. Since our investigation focuses explicitly on the BEC transition, we employ a quantum statistical mechanical framework. The confined Bose gas obeys the Bose-Einstein distribution, $f=1/\{\exp[(E-\mu)/(k_BT)]-1\}$. There is no explicit time variable, and all changes to the system are performed quasistatically to ensure that thermal equilibrium is maintained at all times. Hence, to determine the energy eigenvalues, we numerically solved the time-independent Schrödinger equation, $\hat{H}\Psi=E\Psi$, for the domain $\Omega$. The Hamiltonian of the system includes explicit dependence on the shape parameter, $\theta$,
\begin{equation}
\hat{H}(\theta)=-\frac{\hbar^2}{2m}\nabla^2 + V(x,y,L,\theta).
\end{equation}
The shape dependence enters into the thermodynamic expressions via the energy spectra. The quantum shape effect can be physically understood from the perspective of overlapping quantum boundary layers~\cite{qbl,Aydin2019}, which increases with the rotation angle in these kinds of nested square domains. Thus, increasing angle indicates the increase of the strength of the quantum shape effect.

To make our results independent of the specific type of Bose gas and any other free parameters, we employed dimensionless values throughout this work. Using these initial values, we calculated the discrete energy eigenvalues for each degree step from $\theta=0$ to $\theta=45$ degrees. We use 5000 of them for each case to ensure that the partition function saturates due to the soft cutoff provided by the Bose function. In doing so, we obtain the energy spectrum for each angular configuration. Subsequently, we converted these results into dimensionless values to generalize our findings and eliminate dependence on the chosen initial parameters. 

The dimensionless temperature is defined as $\tilde{T} = T/T_0$, where $T_0$ is the reference temperature obtained from the ground state energy $E_0$ at $\theta=45^{\circ}$ configuration (i.e. minimum $E_0$ value), $T_0=E_0/k_B$. Energies are normalized by $k_BT$ as, $\tilde{E}= E/(k_B T)$, and the dimensionless density is defined as $\tilde{n}_r = n/n_q(T_0)= \lambda_{th}^2(T_0) N/A$, where $n=N/A$ is classical surface density, $N$ is the number of particles and $A$ is the area of the confinement domain $\Omega$. Here, $n_q(T)=1/\lambda_{th}^2(T)$ is the quantum density. The dimensionless density is renormalized using the quantum density at the reference temperature to ensure that the density, and thus the number of particles remain constant for different temperatures. The data and the codes are given in Ref.~\cite{BECshapecode}.

\section{Condensate fraction}
Fig. 1 provides qualitative support for the appearance of condensation with increasing $\theta$, indicated by the increased probability density of the ground state. To analyze the BEC transition due to shape effect quantitatively, we determine the condensate fraction,
\begin{equation}
f_{GS}=\frac{N_{GS}}{N}=\frac{\sum_{E_{GS}} f}{\sum_E f},
\end{equation}
where $N_{GS}$ is the number of particles in the ground state. $\sum_{E_{GS}}$ represents the sum over the (fourfold) degenerate ground state energy levels, whereas $\sum_{E}$ represents the sums over all energy levels. We calculate chemical potential numerically from the particle number equation, $\sum_E f$, as an inverse solution by fixing the particle number and temperature.

Variation of the condensate fraction with shape is shown in Fig. 2, for different temperatures, $\tilde{T}$, and densities, $\tilde{n}_r$. The densities are chosen to ensure that the particle number fluctuations remain small. The condensate fraction consistently increases with decreasing temperature and increasing density, similar to ordinary BEC~\cite{BECbook}. The shape dependence of the condensate fraction, however, exhibits more interesting behaviors. We find that the condensate fraction increases with angle for all temperature and density values, indicating the ability to control BEC transition with shape and inducing the condensation while all other control variables are constant. The overall increase of the condensate fraction with quantum shape effect can be understood from the reduction of the ground state energy with increasing angle~\cite{specsist}, which leads to more particles occupying the ground state, thereby increasing the condensate fraction.

We observe that even a slight increase $\theta$ around $5^{\circ}$ can result in a drastic change in the condensate fraction, which then saturates at approximately $\theta = 15^{\circ}$. The pronounced effect of the $0^{\circ}$-$15^{\circ}$ angular range on the condensate fraction is attributed to the characteristic geometry of the system and its impact on the eigenspectrum. Initially, at $\theta=0^{\circ}$, the domain is a square annulus, which gradually transitions into a configuration with four weakly connected identical triangles, leading to a fourfold degeneracy in the spectrum. The vertex distance at different angles, defined as the perpendicular distance from a vertex of the inner square to the nearest side of the outer square, describes the degree of connection between the triangles. Around $\theta=15^{\circ}$ the vertex distance becomes sufficiently small, causing the local particle density near the vertices to approach zero, effectively establishing the fourfold degeneracy well before the $45^{\circ}$ configuration. This can also be seen in Fig. 4, where the thermal distribution function is plotted against the quantum state variable, revealing the subtle emergence of the fourfold degeneracy even at $5^{\circ}$.

To clearly see the influence of the quantum shape effect, we plot the condensate fraction difference between $\theta=45^{\circ}$ and $\theta=0^{\circ}$ configurations changing with the temperature for various densities, in Fig. 2d. While quantum shape effect increases with increasing particle density leading to easier condensation, the sensitivity of the condensation fraction to the changes in shape parameter decreases. In other words, with increasing particle density, more particles occupy the ground state thereby decreasing the difference between the quantum shape effects of $0^{\circ}$ and $45^{\circ}$ configurations.

While quantum shape effects dominate with decreasing temperature, we observe peaks in the sensitivity (condensate fraction difference) of the quantum shape effects at different temperature values for different densities in Fig. 2d. In Fig. 2a, the condensate fraction shows a less noticeable change for $\tilde{T} = 0.1$ because nearly complete condensation has already been established for all angular configurations. Thus, further decreasing the temperature does not enhance the quantum shape effects, explaining the peaks observed in Fig. 2d. These peaks illustrate the intricate interplay when both density and temperature are varied. Additionally, the peak magnitude of the quantum shape effect on the condensate fraction shifts to higher temperatures with increasing density. $36\%$ difference in the quantum shape effect is obtained at $\tilde{T} = 0.7$ and $\tilde{n}_r = 1$. 

For a practical case, choosing Rb atom confined in the flat-bottom core-shell trap (Fig. 1) with side length $L=10~\si{\micro\metre}$, for $\tilde{n}_r=10$ and $\tilde{T}=2$, the temperature is approximately $T=11~\text{nK}$, and the particle density is $n=1.5\times 10^{12}~\text{m}^{-2}$. The condensation fraction difference becomes $\Delta N_{GS}/N=29\%$ for this particular case. In the case of $\tilde{n}_r=5000$ and $\tilde{T}=500$, these values become $T=2.8~\si{\micro\kelvin}$, $n=7.5\times 10^{14}~\text{m}^{-2}$, and $\Delta N_{GS}/N=17\%$.

\section{Specific heat}
Next, we examine the dimensionless specific heat capacity at constant Weyl size parameters (volume, area, periphery, vertices), denoted by $W$ subscript in Eq. (3), under the change of $\theta$ for dimensionless density $\tilde{n}_r = 1$ and dimensionless temperatures $\tilde{T} = \{0.1, 0.5, 1, 5\}$ as seen in Fig. 3a. The specific heat exhibits distinct and intriguing characteristics that differ significantly from those observed in Bose gases confined in regular flat-bottom geometries or harmonic traps, even in its temperature variation, Fig. 3b.

\begin{equation}
    \frac{c_W}{k_B}=\frac{\sum_{\tilde{E}}\tilde{E}^2f(1+f)}{\sum_{\tilde{E}}f}-\frac{\left[\sum_{\tilde{E}}\tilde{E}f(1+f)\right]^2}{\sum_{\tilde{E}}f(1+f)\sum_{\tilde{E}}f}.
\end{equation}

At higher temperatures, the specific heat increases with increasing $\theta$, similar to the behavior observed in systems following the Boltzmann distribution~\cite{aydinphd}. However, this trend reverses at lower temperatures, where the specific heat decreases with $\theta$. A peak forms at $\tilde{T}=0.5$, but it disappears when the temperature is further reduced to $\tilde{T}=0.1$, for example. At low temperatures, thermal contributions mainly come from low-lying (near-ground) states, where the saturation to ground state diminishes the system's ability to absorb heat, which is more pronounced at higher angles. This situation does not occur at higher temperatures because many thermal states are already available to be occupied, contributing to the specific heat.

Fig. 3b shows the temperature dependence of the specific heat for different angular configurations, allowing for a more direct comparison with the textbook results. Since the system we study has a finite geometry with a finite number of particles, there is no sharp peak in the temperature behavior of the specific heat, indicating a gradual condensate transition. The typical specific heat peak is observed for angular configurations larger than $\theta = 5^{\circ}$. Conversely, a conventionally unexpected dip appears at $\theta = 0^{\circ}$, which is also faintly present at $\theta = 5^{\circ}$. The cause of this dip can be understood by examining the energy spectrum and thermal occupation function, in Fig. 4, where significant differences in the thermal occupation of energy states are observed for various shapes.

At $\theta = 0^{\circ}$, the 2D square annulus domain can also be considered as a 1D bent wire with connected endpoints. This effective reduction in dimensionality is evident in both the energy spectrum and the local features of the eigenfunctions. For clarity, we will refer to the direction along the bent wire as the axial direction and the perpendicular direction to this axis as the lateral direction, corresponding to the width of the wire. Upon plotting the eigenfunctions for the square annulus domain, we observed that before reaching the $E_{42}$ (42nd energy state), the eigenfunctions primarily extend along the axial direction because their wavelengths are too large to fit into the lateral direction. In this regime, $\theta = 0^{\circ}$, the energy states are mainly influenced by the length of the wire, with minimal contributions from the lateral dimension, resulting in a quasi-1D behavior where the energy levels are spaced according to the axial confinement. Starting with $E_{42}$ (the 42nd energy state), the wavelength of the eigenfunctions becomes small enough to fit one full wavelength in the lateral direction of the bent wire (corresponding to the first excited state of the lateral direction), Fig. 4a. This situation is analogous to the opening of subbands in confined systems, hence we denote this phenomenon as the appearance of subbands. The appearance of the second subband can also be seen around $E_{120}$ at $\theta = 0^{\circ}$.

The emergence of these local subbands introduces sharp discontinuities or kinks in the energy spectrum and in the thermal distribution, which are crucial for the behavior of thermodynamic quantities such as specific heat. These subbands cause the dips in the $\theta = 0^{\circ}$ and $\theta = 5^{\circ}$ configurations in specific heat. This highlights the intricate interplay between shape-induced quantum effects and the thermodynamic properties of the system. Note that at $0^{\circ}$ configuration, as the eigenfunctions for higher energy states begin to fit into the lateral direction of the quasi-1D wire, the number of available energy states per unit energy step increases, and the degeneracy almost disappears. The ground state energy is also at its highest at $0^{\circ}$ configuration among all angular configurations. Consequently, this shape exhibits the most resistance to the BEC transition. As $\theta$ increases, the quasi-1D wire characteristic diminishes, Fig. 4b. The subbands start to vanish because the higher wavelengths can start to fit into the thicker parts of the about-to-be-formed triangular regions. In Fig. 4c, at $\theta = 45^{\circ}$, the domain behaves as if it is divided into four separate triangle regions due to the bottleneck created by the inner core's edges. This configuration eliminates the quasi-1D behavior, resulting in the complete disappearance of subbands. Fourfold degeneracy is restored, and the kinks in the thermal distribution disappear. The ground state energy is also reduced with increasing $\theta$, making BEC easier to form.

\section{Conclusion}
In conclusion, we have introduced the phenomenon of shape-induced BEC. Size-invariant shape transformations lead to unique behaviors in the condensate fraction and specific heat, offering additional control over the properties of Bose-Einstein condensates. Our findings reveal that the condensate fraction is significantly influenced by the shape parameter. Moreover, we have observed that the shape-dependent specific heat exhibits markedly different behaviors compared to conventional non-interacting Bose gases confined in other potentials.

The introduction of quantum shape effects in Bose gases allows us to control the sensitivity of the BEC transition with a new degree of freedom: shape. Unusual variations in specific heat due to shape can be harnessed to design highly sensitive electric and magnetic field sensors \cite{wildermuth2006sensing}. Experimentally observing these variations can verify the existence of quantum shape effects on BEC.

We plan to extend this research by further exploring the quantum shape effects on BEC itself. Specifically, we will investigate the influence of inter-particle interactions, examine the effects of finite-time changes with explicit modeling of a heat bath, and explore potential applications of shape-induced control in quantum technologies. Understanding the nature of the quantum shape effect in BEC could have far-reaching consequences in this direction. For example, the ability to control the condensate fraction and specific heat through shape manipulation could lead to the development of new types of quantum sensors. These sensors could leverage the sensitivity of BEC to shape changes, enabling precise measurements of external fields or forces by monitoring variations in the specific heat.

\section{Acknowledgements}
A.A. acknowledges financial support from the Sabanci University Integration Projects Support with project code B.A.CF-24-02877.

\bibliography{refs}
\end{document}